# Emergence biases in molecular evolution

**Timothy Fuqua[1,2,*] and Nikolaos Vakirlis[3]**

1) Institute of Ecology and Evolution, University of Bern, Bern, Switzerland.
2) Swiss Institute of Bioinformatics, Quartier Sorge-Batiment Genopode, Lausanne, Switzerland.
3) Hellenic Pasteur Institute, Athens, Greece.
***For correspondence**: timothy.fuqua "at" unibe.ch

## Abstract

Biases in molecular evolution can significantly influence evolutionary trajectories. They have been described in a variety of contexts such as development and mutation, but not for acquiring new functions (i.e. emergence). Here, we formalize the term, *emergence bias*, as the molecular predisposition that, upon mutation, biases a genetic sequence towards or against gaining new functions or causing new phenotypes. These biases have been observed in previous studies for the emergence of promoters, enhancers, and de novo proteins, but never formally characterized as such. In this Perspective piece, we describe these studies and synthesize their findings through the prism of a unifying term, emergence bias, to provide support for this new concept , and speculate on its molecular underpinnings. We believe that emergence biases may play an important role in evolutionary innovations.

## Significance

The extent to which new biological functions emerge from mutations can substantially differ between DNA sequences, but has never been formally characterized. This work describes examples of such biases in molecular evolution, particularly for the emergence of new regulatory DNA and proteins, and speculates on the mechanisms underlying the biases. The authors propose this phenomenon be described as "emergence biases" in molecular evolution.

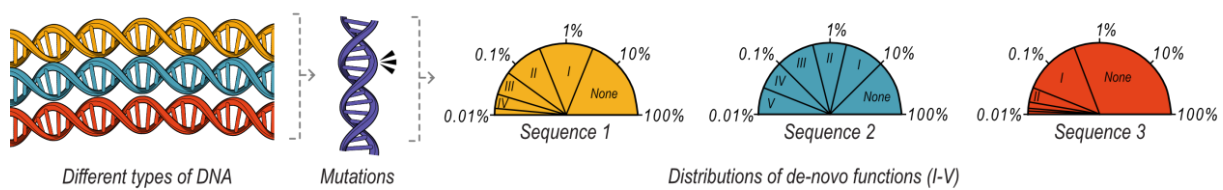

**Figure 1 Emergence biases.** Emergence bias is a molecular predisposition that, upon mutation, biases a genetic sequence towards or against gaining new functions. Left: different genetic sequences colored in yellow, blue, and red. Middle: sequences are exposed to mutations. Right: the frequency of new functions, labeled I-V (log-scale) acquired in the mutants.

## Introduction

There are a number of characterized biases underlying molecular evolution. These include developmental biases where mutations produce an unequal distribution of forms [1], mutational biases where the rates of different types of mutations differ [2], and codon biases where synonymous codons are not uniformly encoded [3]. Each of these biases can significantly influence what phenotypic variation is possible, effectively steering evolution in particular directions (e.g. developmental [4,5], mutational [6–8], codon [9]).

To describe a bias, one must specify what it is relative to. For example, developmental biases are relative to a default assumption that all forms evolve equally within a morphospace [10]. Similarly, codon biases are relative to a null assumption that all synonymous variants are encoded equally [3]. For many cases, however, there is not a clear null or default assumption. When this occurs, biases of one molecule can also be relativized to another molecule. For instance, one could compare the evolutionary potential of a DNA sequence to that of a randomly synthesized DNA sequence [11–13], foreign DNA [14,15], or the reverse version of the sequence of interest [12]. The caveat with such relative comparisons, however, is they cannot preserve all aspects of a tested sequence. For instance, scrambling a DNA sequence may maintain AT-content, but does not maintain dinucleotide composition. Arguably, one of the challenges of characterizing biases in molecular evolution is finding the appropriate null models and assumptions.

Furthermore, the directionality of a bias is described using terms like *drive* [16] or *constraint* [17]. One potential issue with these terminologies is that *drives* connotate with positive selection, while *constraints* with negative selection. Their distinction, however, is arguably redundant, as a drive towards one phenotype is simultaneously a constraint against another [16]. In this Perspective piece, we simply refer to both drives and constraints as *biases*, and specify their directionality as either *towards* or *against* a particular trait. Under this view, directionality of a bias is irrelevant to selection.

Recent developments in computational and experimental approaches have begun to reveal that the emergence of novel functions (i.e. any molecular activity that confers a phenotype) from non-functional sequences, is an important aspect of evolutionary innovation [18]. Examples of emergence have been described from regulatory DNA [19,20], to the emergence of entire protein-coding sequences [21]. To the best of our knowledge, there is no formally described term that can apply to the biases underlying such emergence. The term "preadaptation" [22] and the related term "exaptation" [23] cover cases of traits originally evolved for one function that acquire a new one, but are tightly linked to the action of selection and strongly associated with organismal evolution. This Perspective piece aims to unite similar ideas and concepts throughout the literature under a missing umbrella term: *Emergence bias,* which we define as a molecular predisposition that, upon mutation, biases a genetic sequence towards or against gaining a new function or phenotype (**Fig 1**). Here, we present evidence for emergence biases across different subfields of molecular biology, and discuss the possible mechanisms underlying them.

# Evidence of emergence biases
### Prokaryotic promoter emergence biases
Prokaryotic promoters are DNA sequences that RNA polymerase binds to for initiating transcription [24]. Mutations to non-promoter sequences can readily create new promoters from as little as single point mutations [11,13,25–27]. Furthermore, upon synthesizing randomly generated DNA sequences, approximately ~10% will have promoter activity, depending on their sequence lengths and AT-contents [11,13,28,29].

Because prokaryotic promoters emerge relatively frequently, they present a unique system to explore how emergence differs in various types of sequences. For example, in three independent studies, we isolated non-promoter DNA that was i) randomly synthesized [27], ii) from the *E. coli* genome [27], iii) from AT-rich regions of the *E. coli* genome called "Promoter Islands" [25], or iv) from the ends of a family of pervasive transposons called IS3s [26]. We created random mutagenesis libraries from these different sequences, and screened their respective mutants for novel promoter activity using Sort-Seq [30,31] (**Fig 2A**).

For each mutagenized sequence, we calculated the probability $P_{new}$ that any given (random) mutation will generate de novo promoter activity. These values significantly differ between the categories of DNA (**Fig 2B**, Kruskal-Wallis test, H=88, p=5.98×10$^{-19}$). For example, we find that random DNA is ~3 times more likely to acquire new promoter activity compared to genomic DNA [27] (**Fig 2B**). Thus, relative to randomly synthesized DNA, we argue that the *E. coli* genome has an emergence bias *against* creating new promoters, or conversely, that random DNA has an emergence bias *towards* creating new promoters, relative to the *E. coli* genome.

We also compared promoter emergence within the sequence categories. For example, the promoter island $P_{new}$ values range from as low as 0.2% to as high as 41% [25] (**Fig 2B**). For the mutagenized pieces of mobile DNA without promoter activity, these probabilities also range extensively ($P_{new}$ = 2-23%) [26] (**Fig 2B**). These studies demonstrate that relative to each other, groups of similar sequences can also vary in their propensity to create de novo promoters [25–27]. We propose that these observations also be described as *emergence biases* relative to their categories (**Fig 2C**).

### Eukaryotic enhancer emergence biases
Enhancers are eukaryotic DNA sequences which control the spatial-temporal patterning of gene expression [40], and encode binding sites for proteins (transcription factors, TFs) to bind and activate or repress transcription [41]. Multiple studies demonstrate how de novo enhancers can emerge from point mutations [19,20,42], allowing us to also observe emergence biases of enhancers.

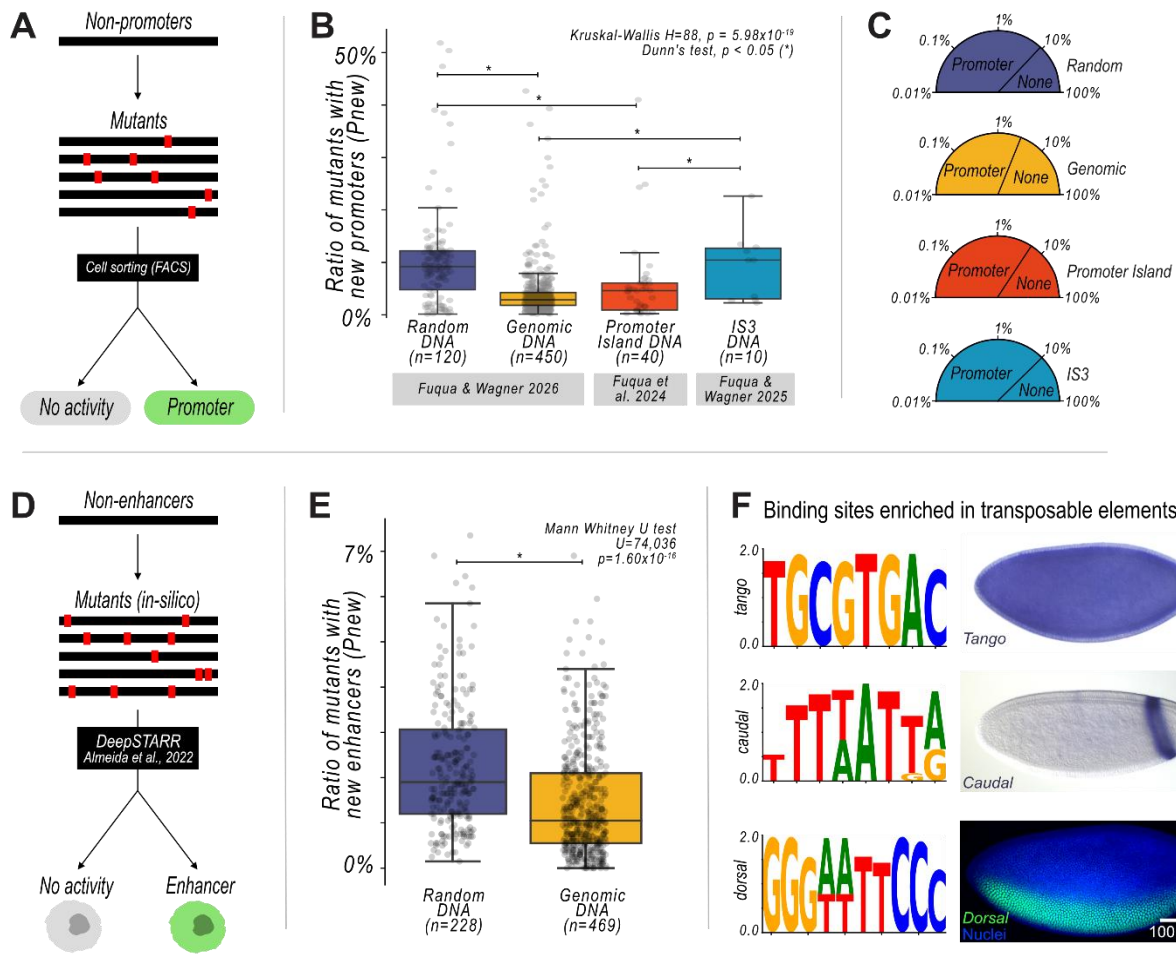

**Figure 2 Emergence biases of cis-regulatory DNA. (A)** Previous works experimentally mutagenize different categories of non-promoter DNA sequences, and test the mutants for promoter activity in *E. coli*. **(B)** The ratio of mutants with new promoters ($P_{new}$) for each mutagenized sequence (individual points). Each box plot corresponds to a category of non-promoter DNA from the studies listed below. Data were all acquired using the same method sorting and gating strategy (Sort-seq). We test the null hypothesis that the $P_{new}$ distributions are the same using a Kruskal-Wallis test (H=88, p=5.98×10$^{-19}$) and use the Dunn's post-hoc test to identify significant differences between the categories. Horizontal lines with asterisks (*) above indicate a p-value < 0.05. **(C)** We plot the median $P_{new}$ value for each category. **(D)** We randomly mutagenized DNA sequences not predicted to have enhancer activity in *Drosophila* S2 cells, and test the mutants for new enhancer activity in-silico using the DeepSTARR sequence-to-function model [32]. **(E)** The ratio of mutants predicted to be enhancers ($P_{new}$) for each mutagenized sequence (individual points) using DeepSTARR. Each box plot corresponds to a category of non-promoter DNA (left: randomly generated DNA, right: fragments of the *Drosophila melanogaster* genome). We test the null hypothesis that the distributions are the same using a Mann-Whitney U test (U=74,036, p=1.60×10$^{-16}$). See **Methods** for simulation details. **(F)** Left: sequence logos for three transcription factor binding motifs enriched in *Drosophila* transposable elements identified in ref [33]. Logos are derived from the Fly Factor Survey database [34] and drawn using Logomaker [35]. Right: representative gene expression patterns for *Drosophila melanogaster* embryos at approximately developmental stage 5. Images for Tango and Caudal are in-situ hybridizations for the respective transcripts from the Berkeley Drosophila Genome Project (BDGP) [36–38]. The Dorsal embryo was stained using the 7A4 antibody [39] from the Developmental Studies Hybridoma Bank, and imaged using a confocal microscope (see **Methods**).

Similar to bacterial promoters, the DNA sequence of a non-enhancer also appears to bias enhancer emergence. For example, Taskiran et al. recently used a sequence-to-function model called *DeepFlyBrain* [43] to predict enhancer activity in *Drosophila* KC cells from given DNA sequences [44]. The authors then created stepwise permutations in-silico to non-enhancer sequences which incrementally improved their predicted scores (i.e. greedy walk). The model predicted that some sequences required just 6 point mutations to become "from scratch" KC enhancers, while others required up to 15 point mutations, and still had weak predicted strengths [44]. While there are still limitations to predicting functions from point mutations using sequence-to-function models [45,46], these simulations suggest that relative to each other, the former sequences have an emergence bias *towards* becoming enhancers, and the latter *against* becoming enhancers.

Here, to further explore the emergence biases of enhancers, we used a similar sequence-to-function model called *DeepSTARR,* which predicts enhancer activity in *Drosophila* S2 cells [32]. Using DeepSTARR, we first identified 228 non-enhancer sequences from randomly generated DNA, and 469 non-enhancer sequences from the *Drosophila melanogaster* genome. For each of these starting sequences, we quantified the likelihood of random mutants acquiring new enhancer activity (**Fig 1D**) (1-10 point mutations per mutant; 2,000 mutants per starting sequence, see **Methods**). We find differences in emergence likelihoods within the random ($P_{new}$ = 0.15 - 7.4%) and genomic ($P_{new}$ = 0.0 - 6.9%) starting sequences, as well as between the sequence groups, with enhancers emerging nearly twice as frequently from mutants in random DNA vs genomic DNA (median $P_{new}$ = 1.9% vs 1.0%, Mann-Whitney U test, U=74,036, p=1.60×10$^{-16}$) (**Fig 1E**). These simulations suggest once again, that there are biases underlying enhancer emergence.

In another study, we synthesized random libraries of DNA, and screened them for enhancer activity in *Drosophila* embryos [47]. We found that the majority of sequences had detectable expression in the embryos, but their expression patterns were biased based on which TF binding sites the random DNA serendipitously encoded. This finding resonates with a recent study, which demonstrates that in *Drosophila,* de novo genes (i.e. those originating from non-coding DNA) are more likely to be regulated by a subset of specific TFs (e.g. Vis, Achi, and Jri), and by extension, expressed in specific cell and tissue types (e.g. testis and accessory glands) [48].

These findings also reflect related studies that transposable elements (TEs) are biased towards recruiting specific TFs, and can thus bias the emergence of tissue or response-specific gene expression (**Fig 1F**) [33,49–52]. For example, three pervasive *Drosophila* TE families *1360, Cr1a*, and *roo* are each significantly enriched with binding sites for Tango [33,51], which plays an important role in hypoxia response [53]. This finding suggests that *Drosophila* TEs may have an emergence bias towards creating hypoxia-response elements. The *roo* elements are also enriched with binding sites for Dorsal – an early developmental TF - and Jockey elements are enriched with binding sites for Caudal, another developmental TF [33]. These TFs are differentially expressed throughout development (**Fig 1F**). Based on these findings, we speculate that TE-based enhancers may be biased towards emerging within these domains.

Overall, the aforementioned studies demonstrate a proclivity of individual sequences or groups thereof, to become enhancers relative to others. They additionally demonstrate a biased recruitment of specific TFs in de novo enhancers, controlling the tissues and cell types in which de novo enhancers are active. We propose describing this proclivity for different DNA sequences to become enhancers, and drive expression in particular cell types, as emergence bias.

### *Protein emergence biases*
Novel protein-coding genes can evolve from previously non-coding sequences through a process known as de novo gene birth [21]. Once considered wholly implausible, the existence of de novo genes has now been demonstrated in organisms across all domains of life, from plants [54], to fungi [55], to human [56], and bacteria [57]. A growing body of evidence suggests that pervasively translated small Open Reading Frames (sORFs) - which naturally occur in different types of transcripts - provide the substrate for de novo gene birth [58] and serve as the foundation from which a proto-gene can evolve [59].

Such evolutionarily nascent, translated sORFs could have an emergence bias by being "preadapted", that is, already exposed to selective pressures. Studies have predicted that as lowly expressed polypeptides are exposed to selection, deleterious ones (e.g. those prone to aggregate [60]) should be purged from the genome as soon as they appear [22]. Hence, the pool of expressed polypeptides would be enriched for those with a higher chance of evolving into a functional protein because they will not be toxic and in turn, result in evolutionarily young genes having exaggerated gene-like properties. Consistently, low expression is an almost universal feature of young genes [61] and young mouse and *Drosophila* proteins have more intrinsic structural disorder (ISD) than both older ones and random intergenic ORF-encoded polypeptides [62,63]. However such trends are not found in all species (e.g. the opposite ISD trend is found in yeasts [55,64], studies have reported no trends over time [65]) and this difference in trends is partly explained by genomic GC-content [64].

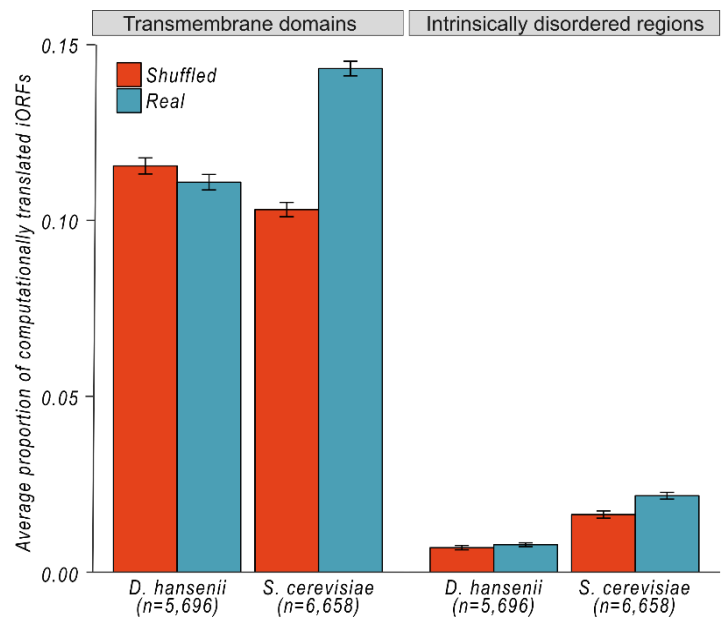

**Figure 3 protein level biases in intergenic regions.** Average proportions of iORFs, full intergenic regions computationally translated into proteins after removal of in-frame stop codons, and their shuffled counterparts (see **Methods** and Tassios et al. [69]), in two species of budding yeasts, that constitute transmembrane domains as predicted by Phobius and intrinsically disordered regions as predicted by IUPRED3.

Emergence biases could also work to prevent the aberrant expression of potentially damaging polypeptides. A recently identified mechanism relies on the presence of a C-terminal hydrophobic tail, which targets polypeptides for proteasomal degradation through a pathway that is also used for membrane targeting [66]. This property is abundant in polypeptides encoded by sORFs outside of known protein-coding regions but depleted within them. Another mechanism relies on the presence of degrons, short terminal motifs that target proteins for proteasomal degradation, and they also appear to be more prevalent within polypeptides encoded by noncanonical sORFs than canonical protein-coding genes or randomized controls [67].

The sequence of an sORF may bias the functions it can acquire. In a previous study, we showed that intergenic regions of the *S. cerevisiae* genome have a surprisingly high potential to, if theoretically translated, encode polypeptides with computationally predicted transmembrane (TM) domains [68]. Follow-up studies reveal this bias also applies to the majority of *Saccharomycotina* yeasts [69] and to a lesser extent to *Enterobacteriaceae* bacteria [70]. In **Fig 3** (left) we show an example comparison of *S. cerevisiae*, that shows the bias, to *Debaryomyces hansenii*, a species that doesn't. Our most recent work [71] revealed that this bias is caused by the existence of polyA/T tracts, which are abundant and enriched in intergenic regions, and are known to function as Nucleosome Depleted Regions (NDRs) to enable transcription of surrounding genes. At the same time, It is possible that these polyA/T tracts are part of an unknown translation mitigation mechanism [72] like the one recognizing hydrophobic CTDs [66]. Regardless, the presence of polyA/T tracts likely biases de novo proteins to be initially localized within specific intracellular locations and encode specific biochemical properties. Consistently, a previous study has shown that *D. hansenii* has much shorter polyA/T tracts which do not appear to be able to repulse nucleosomes [73], possibly an adaptation to the high salt concentration environments that this species inhabits. In contrast to transmembrane domains which are predicted to be prevalent in intergenic regions of both species, intrinsically disordered segments are predicted to be much rarer in both genomes and hence, while a weak enrichment can be seen in *S. cerevisiae* relative to shuffled controls, it is essentially inconsequential (**Fig 3**, right).

# Explanations for emergence biases

Many cases of emergence biases can be explained by the topology and density of the genotype-function space. To illustrate this, we represent all possible DNA sequences within a high-dimensional genotype space, which we illustrate in **Fig 4A** using two dimensions (for simplicity). Here, the closer two sequences are to each other, the more similar they are. Within this genotype space exist different clusters corresponding to unique biological functions. Depending on the given DNA sequence, some sequences will lie closer to a subset of functions vs others. For example (see **Fig 4A**), the yellow genotype lies closest to the hypothetical Function I vs Functions II-V. In this example, point mutations in the yellow genotype will explore the surrounding genotype space, and a subset of these point mutants will overlap with (and thus acquire) Function I. In this case, the yellow genotype has an emergence bias towards Function I.

While this genotype-function map is (and will likely remain) largely hypothetical, it can be useful to understand why some sequences have an emergence bias towards or against acquiring particular functions. For instance, DNA sequences are more likely to become promoters if they contain motifs or signatures which moderately resemble canonical promoters [26,27]. They are also more likely to become promoters as their AT-content increases [27,29]. Because of this resemblance, such sequences would lie close to the boundary of the "promoter function" space in the map, just like the yellow genotype is to Function I (see **Fig 4A**). As another example, transposon coding sequences can coincidentally encode binding sites for particular TFs [33,51] (see **Fig 2F**). Encoding such sites decreases the Euclidean distance between the transposon and the boundaries of the "enhancer" or "promoter" functional space.

The bias towards encoding transmembrane (TM) domains within yeast intergenic regions can also be explained as protein-level functionality mapping to many different areas of the genotype space, i.e. to sequences of low information. This is because highly repetitive polyA/T tracts, and the respective amino acids they encode (Phe, Lys, Tyr), enable the formation of TM helices largely due to their hydrophobicity [68,74] and not their specific amino acid sequences. Because the intergenic regions of the yeast genome are enriched with repetitive sequences of A's and T's [71] - whose primary function might be to hinder nucleosome binding [75] - intergenic regions would also lie close to the transmembrane domain function space. Note: these motifs are removed relatively quickly during the evolutionary time following the initial emergence of a gene and are extremely rare in conserved genes, including those encoding TM domains [71]. Thus, the low information content of these motifs is only relevant to their role in the emergence of novel proteins and is not a feature of TM proteins more generally.

Within the same space, some functions can be, by their very nature, more easily accessible by different sequences. For example, promoters and enhancers operate through protein binding sites for RNA polymerases and transcription factors (TFs). The *information content* of these binding sites differs for each factor, where some TFs bind to very specific sequences (high information), while others to a variety of degenerate sequences (low information) [76–78]. In other words, protein binding "specificity" influences which TFs regulate a given DNA sequence [78]. In truly random DNA, this information content is directly proportional to the probability that that TF will bind a given sequence [27,47,78]. The differences in the information content of TF binding may also explain why de novo enhancers and promoters use particular subsets of (low-information) TFs [27,47], and why de novo genes recruit less TFs compared to older genes [48,79].

Here, to further explore how information content influences enhancer emergence, we acquired a set of 225 TF binding motifs for *Drosophila melanogaster* [80] and calculated their respective information contents (**Fig 4B**), which range from 7.3 – 24.8 bits (median 11.6 bits), similar to previous reports for other eukaryotic TF binding motifs [78]. We classified TFs with less than 11.6 bits (50$^{th}$ percentile) as "low information" TFs, which will bind more frequently to DNA sequences. These low information TFs are thus more likely to be recruited for de novo enhancer emergence. See **Data S1** for the TFs and their respective information contents.

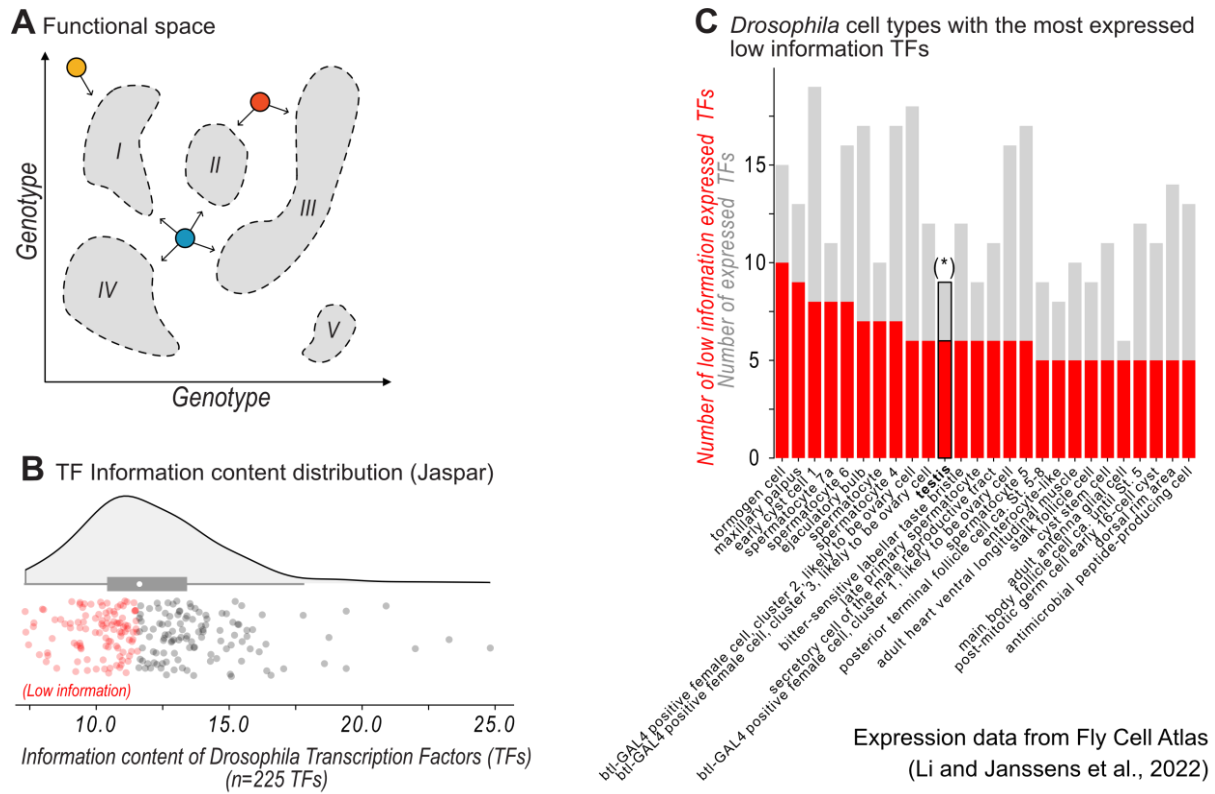

**Figure 4. Information theory and sequence space influence emergence.** **(A)** A hypothetical representation of different functions (dashed gray shapes labeled I-V) in genotypic space (x and y-axes). The colored circles correspond to the three different DNA sequences in the genotypic space. The arrows show which functions these sequences are biased towards based on their proximity in the space. **(B)** The information content of 225 transcription factors (TFs) in *Drosophila melanogaster* from the JASPAR database [80]. TFs below the median are classified as "low information" TFs and are labeled in red. See **Data S1** for the identity of each TF and its respective information content. **(C)** Bar plot of different *Drosophila* cell types from the Fly Cell Atlas [81] and the number TFs (from panel B) expressed in each. See **Data S2** for the remaining cell types.

Because TFs are also differentially expressed, we can extrapolate from the information content to estimate in which cell types, information theory predicts new enhancers are more likely to emerge from. To this end, we used the Fly Cell Atlas, which contains single-cell RNA-seq expression data for 246 cell types in *Drosophila melanogaster* [81]. We used TF transcript levels to infer whether or not each of the 225 TFs are expressed in each cell type, and then counted how many of the low-information TFs are expressed in each cell type (see **Methods**). In **Fig 4C**, we plot the cell types which express at least five low-information TFs (see **Data S2** for the complete set of counts per cell type). Thus, based on information theory and these data, we speculate that new enhancers may be biased towards producing expression specifically in these cell types compared to other cell types. They include germline cells (e.g., spermatocytes, oocytes), somatic support cells, and various epithelial and secretory cell types. Interestingly, the testis are also part of this subset, which could potentially explain why so many de novo genes are expressed in the *Drosophila* testis [48].

# Future directions and conclusions

Here, we provide evidence for emergence biases in the context of de novo promoters, enhancers, and proteins. Such biases dictate the likelihood of different sequences acquiring these functions relative to each other or relative to random DNA. They appear to be governed by the information content encoded within these functions, and the space a sequence occupies. We believe that a critical question for future investigations is whether such biases are more commonly the result of selective filtering or mutational biases, either for some specific functionality or because they make genomes more evolvable, which can be beneficial over an unbiased "naive" state.

We see no reason why emergence biases would be limited to promoters, enhancers, and transmembrane domain functions, nor assume that there are no other mechanisms underlying them. Like all biases in molecular biology, we can envision how emergence biases can strongly influence which phenotypes mutations can produce, potentially steering evolution in particular directions. However, we do not know the extent to which emergence biases can impact evolution at this time. We encourage others to think about these biases and to build upon the concept.

# Methods

### *Data availability*
We provide the data and scripts used to create the figures and analyses in this study at the GitHub repository: https://github.com/tfuqua95/emergence_bias .

### *DeepSTARR enhancer predictions for Drosophila S2 cells*
To calculate enhancer $P_{new}$ values for the different categories of DNA, we first isolated non-enhancer "parent" sequences. For the random DNA parents, we randomly generated a library of 5,000 DNA sequences, each 249 bps in length *in-silico*. From each of these candidate sequences, we used the following DeepSTARR command to predict the developmental enhancer activity of the 5,000 sequences:

```
python DeepSTARR_pred_new_sequence.py -s query_sequences.fasta -m DeepSTARR.model
```

From the predictions, we selected 228 parents from the randomly generated DNA sequences with no predicted developmental enhancer activity, by selecting sequences ranging with prediction scores from -0.1 to +0.1 RNA log2 fold-change (FC) levels. For each parent, we then computationally generated 2,000 unique mutants with 1-10 random point mutation each. We then predicted the developmental enhancer activity for each of these mutant "daughter" sequences using the previous DeepSTARR command, and calculated the parent's respective $P_{new}$ value by dividing the number of daughters with predicted enhancer activity (log2 FC values > +1.0) by the total number of daughters (2,000 daughters each).

To isolate genomic sequences without enhancer activity, we randomly selected 5,000 genomic sequences from the dm6 assembly [82], sampling evenly and exclusively within chromosomes 2L, 2R, 3L, 3R, X, and Y. We identified 469 genomic sequences with log2FC levels between -0.1 to 0.1 using the previous DeepSTARR command, and subsequently analyzed respective mutagenesis libraries as described above.

### **Identifying cell types expressing low information content transcription factors**
We acquired a set of 290 *Drosophila* transcription factor (TF) binding matrices from the JASPAR database [80] (core insecta, non-redundant PFMs) using the following link: https://jaspar.elixir.no/downloads/ . If multiple matrices existed for a given TF, we chose the matrix with the highest information content. We acquired the TF expression data for *Drosophila* cell types from Table S3 of the Fly Cell Atlas [81]. The data contained 246 unique cell types with 501 unique TFs.

We filtered the Fly Cell Atlas dataset to only focus on TFs present in both datasets (n=225 TFs). See **Table S1** for the information content values (in bits) for each of the 225 TFs. To define a TF as "expressed" within each cell type, we noted if it appeared within any of the three columns of the supplemental Table S3 from Fly Cell Atlas: 'high_genes (expr ratio > 50%)', 'medium_genes (50% > expr ratio > 5%)', or 'low_genes (expr ratio < 5%)'. With this approach, a TF is either present or absent in each cell type, and we did not consider transcript dosage.

For each cell type, we then counted how many of the 225 shared TFs were present in each cell type (see **Fig 4C**, gray bars). We then isolated a subset of the TFs below the 50[th] percentile of all information content values, which we call the "low information TFs" and repeated the counting procedure for each cell type (see **Fig 4C**, red bars). **Fig 4C** only displays a subset of the data. See **Table S2** for the data on all cell types.

### **Dorsal antibody staining**
We added a strain of w[1118] *Drosophila melanogaster* lines to an egg collection chamber and collected embryos after an overnight incubation. We washed the collected embryos with saline, and dechorionated them using a 50% bleach solution for 90 seconds, followed by washing with water. We then transferred the embryos to scintillation vials and fixed the embryos with 700 μl 16% PFA, 1.7 ml PBS/EGTA, and 3.0 ml 100% heptane for 25 minutes while shaking at 250 rpm. After fixation, we removed the lower phase of the solution and added 100% methanol to carry out an isotonic shock, followed by subsequent vortexing. We removed embryos from the interphase and upper phases of the vial.

We stained the embryos using an anti-dorsal antibody (7A4, DSHB), and conjugated the antibody with a secondary AlexaFluor antibody (1:500, Invitrogen). The anti-dorsal 7A4 antibody [39] was deposited to the DSHB by Steward, R. (DSHB Hybridoma Product anti-dorsal 7A4). We mounted the fixed and stained embryos using Prolong Gold (Thermo Fischer Scientific) with DAPI, and imaged the embryos using a Zeiss 880 confocal microscope (Zeiss, Germany).

**Protein-level biases in iORFs**

Proportions of amino acids predicted to be in transmembrane protein domains in real and single-nucleotide shuffled iORFs in *S. cerevisiae* and *Debaryomyces hansenii* were obtained from Tassios et al. [69]. The same iORF sequences were given as input to IUPRED3 [83] with the *-long* parameter to obtain the numbers of amino acids predicted to belong in intrinsically disordered regions (with probability above 0.5). For each sequence, this number was divided by the sequence length to obtain the proportion.